\documentclass[aps,prb,reprint, amsmath, amssymb]{revtex4-1}

\usepackage{bm}
\usepackage[retainorgcmds]{IEEEtrantools}
\usepackage{epsfig}

\usepackage{graphicx}
\usepackage{subfigure}%
\usepackage{mathrsfs}
\usepackage{amsmath}
\usepackage{amssymb}
\usepackage{color}
\usepackage{amsfonts}

\newcommand{\ud}{\,\mathrm{d}}

\DeclareMathOperator{\tr}{tr}

\newcommand{\trans}{^{\text{T}}}

\begin{document}

\title{The open Heisenberg chain under boundary fields: a magnonic logic gate}
\date{\today}
\author{Gabriel T. Landi}
\email{gtlandi@gmail.com}
\affiliation{Universidade Federal do ABC,  09210-580 Santo Andr\'e, Brazil}

\author{Dragi Karevski}
\affiliation{Institut Jean Lamour, Department P2M, Groupe de Physique Statistique, Universit\'e de Lorraine, CNRS, B.P. 70239, F-54506 Vandoeuvre les Nancy Cedex, France}

\begin{abstract}

We study the spin transport in the  quantum Heisenberg spin chain subject to boundary magnetic fields and driven out of equilibrium by  Lindblad dissipators. 
An exact solution is given in terms of matrix product states, which allows us to calculate exactly the spin current for any chain size. 
It is found that the system undergoes a discontinuous spin-valve-like quantum phase transition from ballistic to sub-diffusive spin current, depending on the value of the boundary fields. Thus, the chain behaves as an extremely sensitive magnonic logic gate operating with the boundary fields as the base element.
\end{abstract}
\maketitle{}

%
%
%
%

\section{\label{sec:int}Introduction}

One of the fundamental issues in condensed matter physics is the determination of macroscopic parameters from the underlying microscopic properties. 
For systems in equilibrium, the Gibbsian approach gives an elegant solution since it depends only on the underlying microscopic energy spectrum. 
However, even if substantial progress has recently been made in understanding non-equilibrium systems, in particular through the so called fluctuation theorems, \cite{Evans1993,*Evans1994,Gallavotti1995b,*Gallavotti1995,Jarzynski1997,*Jarzynski1997a,*Jarzynski2000,Crooks1998,*Crooks2000,Talkner2007,*Talkner2009,*Campisi2011} no such  approach  is available for systems  in a Non-Equilibrium Steady-State (NESS),  characterized by the existence of steady currents. 
This forces one to resort to a full dynamical calculation in order to extract steady-state parameters.
Such a difficulty is inherent of  non-equilibrium systems, dating back to Drude's calculation of the electrical conductivity of metals in 1900. \cite{Drude1900a}
As another example, we note the recent discussions concerning the microscopic derivation of Fourier's law in insulating crystals.
\cite{Rieder1967,Bolsterli1970,Manzano2012,*Asadian2013,Landi2013a,*Landi2014a,Karevski2009,*Platini2010}

A more thorough understanding of the NESS is also essential for the development of several applications in
 phononics, 
 \cite{Terraneo2002,*Li2004a,Chang2006,Landi2014b}
spintronics,  
\cite{Wolf2001,Murakami2003,Fabian2004,Oltscher2014a}
and magnonics. 
\cite{Serga2010,Chumak2014}
We point in particular to two recent remarkable papers by Chumak \emph{et.~al.}\cite{Chumak2014} and Oltscher \emph{et.~al.}\cite{Oltscher2014a}.
In Ref.~\onlinecite{Chumak2014} the authors report on a magnonic logic gate, where the magnon current is adjusted by controlling the number of magnon scattering processes  induced by an auxiliary magnon injector (the base).
On a different setting the authors in Ref.~\onlinecite{Oltscher2014a} study the transport of spin polarized current in a two dimensional electron gas. 
They observe for the first time the existence of a ballistic spin flow, in stark disagreement with classical predictions.

The transport properties reported in Refs.~\onlinecite{Chumak2014,Oltscher2014a} both involve the presence of a NESS. Moreover, they share in common the fact that they cannot be explained by classical theories, thus requiring a full quantum treatment. 
On the theoretical side, these quantum NESS are usually implemented on 1d lattice spin systems coupled to external reservoirs. 
\cite{Bandyopadhyay2011,Benenti2009,Znidaric2011,*Znidaric2013,Prosen2012,Mendoza-Arenas2013,*Mendoza-Arenas2013a,Prosen2011b,*Prosen2011,Karevski2013,Popkov2013a,Popkov2013b,Landi2014b,Yan2009,*Zhang2009} The effect of the reservoirs is quite often described by a non-unitary Lindblad dynamical equation. \cite{Lindblad1976,Breuer2007}
However, these models, being quantum many-body problems,  can seldom be solved exactly and from a numerical point of view they can usually only be solved for small lattices.
 
The purpose of this paper is to study the transport properties in the NESS of the one-dimensional Heisenberg chain coupled to two Lindblad reservoirs at each end, and also subject to magnetic fields at its boundaries. 
Remarkably,  the steady state of this model is exactly expressible in terms of a matrix product state\cite{Prosen2011b,*Prosen2011,Karevski2013} involving operators satisfying the SU(2) algebra (in the case of an XXZ chain this generalizes to the quantum  $\text{U}_q$[SU(2)] algebra\cite{Karevski2013}). 
This provides a method to compute the steady-state spin current $J$ for \emph{any chain size}.\cite{Popkov2013a}
We will show that depending on the strength of the applied magnetic field, $J$ may undergo a discontinuous spin-valve-like   quantum phase transition from  ballistic to sub-diffusive ($J\sim 1/N^2$; cf. Fig.~\ref{fig:h}(d) below). 
As we shall discuss, the origin of this transition is related to the entrapment  of magnons inside the chain caused by the boundary fields which,   in turn, increase the number of magnon scattering events. 
We argue that our system may be used as an extremely sensitive magnonic logic gate operating with an external magnetic field as the base element.

%
%
%
%

\section{\label{sec:mod}Description of the model}

We consider the isotropic  Heisenberg spin-1/2 chain with $N$ sites  described by the Hamiltonian
\begin{equation}\label{H}
H = \frac{1}{2} \sum\limits_{i = 1}^{N-1} \left( \sigma_i^x \sigma_{i+1}^x + \sigma_i^y \sigma_{i+1}^y + \sigma_i^z \sigma_{i+1}^z \right) + h (\sigma_1^z - \sigma_N^v),
\end{equation}
where the $\sigma$'s are the usual Pauli matrices. 
The last term describes the Zeeman interaction experienced by the boundary spins with 
a field pointing in  the $z$ direction on the first site and in the $-\bm{n}_v=(\sin\theta,0,-\cos\theta)$ direction on the last site.
Note that with this parametrization the boundary fields point in opposite directions when $\theta = 0$.

The chain is coupled  to two reservoirs at each end such that its density matrix $\rho$  is  governed by the Lindblad master equation\cite{Breuer2007}
\begin{equation}\label{lindblad}
\frac{\ud \rho}{\ud t} = -i [H,\rho] + D_L(\rho) + D_R(\rho),
\end{equation}
where the left and right dissipators $D_{L(R)}$ are given by 
\begin{equation}\label{DL1}
D_\alpha(\rho) = \sum\limits_{r=\pm} 2 K^\alpha_r \rho{ K^\alpha_r}^\dagger - \{ {K^\alpha_r}^\dagger K^\alpha_r, \rho\},
\end{equation}
with $K^L_\pm = \sqrt{\gamma(1\pm f)} \sigma_1^\pm$  and  $K^R_\pm = \sqrt{\gamma(1\mp f)} {\sigma_N^v}^\pm$ and where 
${\sigma_N^v}^\pm$ are the ladder operators in the  $\bm{n}_v$ direction. Explicitly one has  ${\sigma_N^v}^- = (\cos\theta \sigma_N^x -i \sigma_N^y + \sin\theta \sigma_N^z)/2$ for the lowering operator and the adjoint expression for the raising one. 

\begin{figure}[!h]
\centering
\includegraphics[width=0.3\textwidth]{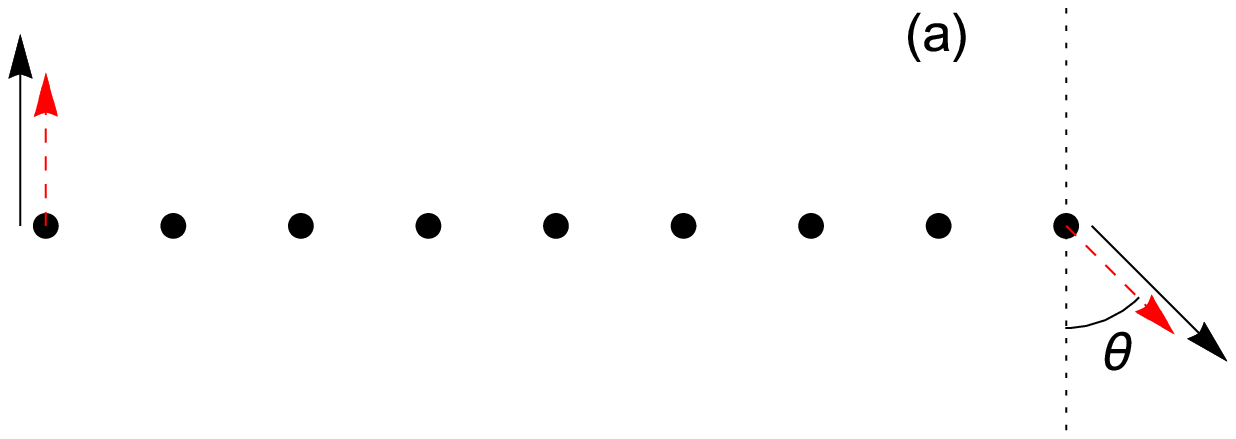}\\
\includegraphics[width=0.3\textwidth]{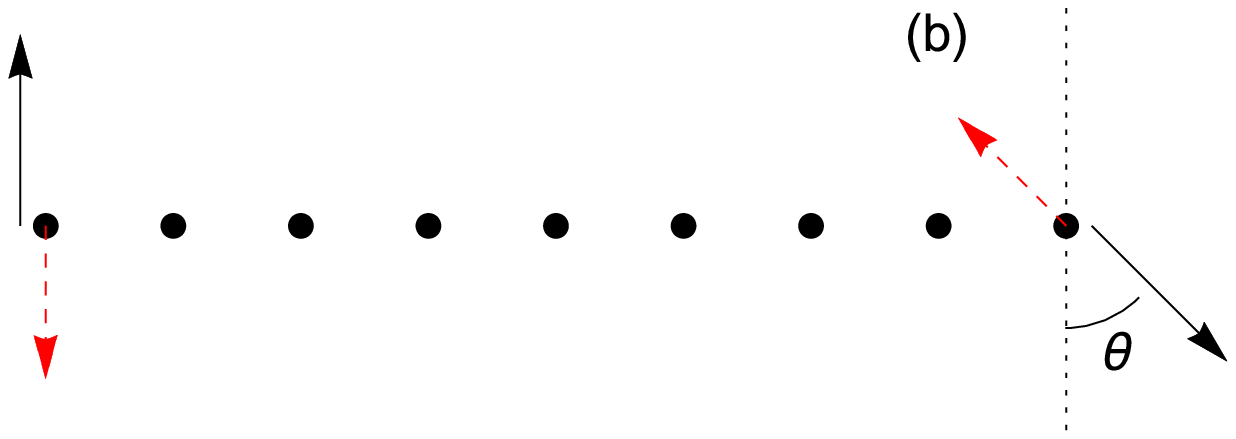}
\caption{\label{fig:drawing}(Color Online) Schematic drawing of the dissipators $D_{L,R}(\rho)$ (black, solid arrows) and the boundary fields $h$ (red, dashed arrows) acting on the first and last spins of the chain. In (a) the fields are in the same direction as the dissipators ($h>0$) and in (b) the fields act in directions opposite to the dissipators ($h<0$).}
\end{figure}

The forcing term $f\in[0,1]$ describes the polarization of the spin reservoirs and is related to a reservoir inverse temperature $\beta$ by $f =\tanh(\beta)$. 
At  $f=1$ (zero temperature) the left bath corresponds to a perfect magnon source, pumping magnons into the system at a rate $\gamma$, while  the right dissipator is a perfect drain, absorbing magnons at the same rate.
We shall concentrate mostly on $f=1$, even though some words will be given for $f<1$.
Note that when $h>0\; (<0)$ the boundary fields point in the same (opposite) direction as the dissipators (Fig.~\ref{fig:drawing}).

%
%
%
%

\section{\label{sec:sol}Outline of the matrix product state solution}

The unique NESS attained by the system at long times is the solution of Eq.~(\ref{lindblad}) with $\ud \rho/\ud t = 0$:
\begin{equation}\label{SS}
i [H,\rho] = D_L(\rho) + D_R(\rho)\; .
\end{equation}
At $f=1$ the exact solution was found in  Ref.~\onlinecite{Karevski2013} in terms of a matrix product state (MPA), as we now outline.
The first step is to note that since $\rho$ is a Hermitian positive semi-definite operator, we may use the following parameterization:
\begin{equation}\label{SSd}
\rho = \frac{SS^\dagger}{\tr (SS^\dagger)}\; .
\end{equation}
For a Heisenberg chain made of $N$ spins 1/2, the operator $S$ lives on the Hilbert space  $\mathfrak{H} = \mathbb{C}^{2N}$. 

We now use the ansatz that $S$ can be described by a matrix-product state:
\begin{equation}\label{S}
S = \langle \phi | \Omega^{\otimes N} | \psi \rangle
\end{equation}
where $\Omega$ is a $2\times 2$  matrix with operator-valued entries
\begin{equation}\label{omega}
\Omega = \begin{pmatrix}
S_z & S_+ \\[0.2cm]
S_- & -S_z 
\end{pmatrix} = S_z \sigma^z + S_+ \sigma^+ + S_- \sigma^-\; .
\end{equation}
The operators $S_a$ live in an auxiliary space $\mathfrak{A}$ so that $\Omega^{\otimes N} \in \mathfrak{H}\otimes \mathfrak{A}$. 
After contracting with $|\phi\rangle$ and $|\psi\rangle$ we recover $S \in \mathfrak{H}$.
From the bulk structure of the Hamiltonian~(\ref{H}), it can be shown that if Eq.~(\ref{S}) is to be a solution, then the operators $S_a$ must obey the SU(2) algebra: 
\begin{IEEEeqnarray}{rCl}
\label{alg1} [S_z, S_\pm] &=& \pm S_\pm  \\[0.2cm]
\label{alg2} [S_+,S_-] &=& 2 S_z\; .
\end{IEEEeqnarray}
The proper representation of the algebra to be explicitly used in the MPS solution is specified by  
a complex representation parameter $p$ which is fixed by substituting Eq.~(\ref{S}) and (\ref{SSd}) into the steady-state equation~(\ref{SS}) and solving the resulting equations. It turns out that it is fixed by a lowest weight condition:  $\langle \phi | S_z=p\langle \phi |$, where $\langle \phi| \equiv \langle 0|$ is a lowest weight state of the representation. Explicitly, in terms of a semi-infinite set of states $\{|n\rangle \}_{n=0}^\infty$ one has the irreducible representations
\begin{IEEEeqnarray}{rCl}
\label{Sz} S_z &=& \sum\limits_{n=0}^\infty (p-n) |n \rangle\langle n | \\[0.2cm]
\label{Sp} S_+ &=& \sum\limits_{n=0}^\infty (n+1) |n \rangle\langle n+1 | \\[0.2cm]
\label{Sm} S_- &=& \sum\limits_{n=0}^\infty (2p-n) |n+1 \rangle\langle n | \; .
\end{IEEEeqnarray}
Notice that for half-integer values of $p$ these representations reduce to the usual finite dimensional representations of SU(2). 
In the present case, the representation parameter turns out to be
\begin{equation}\label{p}
p = \frac{i}{2(\gamma - i h)}
\end{equation}
which fixes the associated infinite dimensional representation of SU(2). The right state $|\psi \rangle$ over which $\Omega^{\otimes N}$ is evaluated is given by the coherent state
\begin{equation}\label{psi}
|\psi \rangle = \sum\limits_{n=0}^\infty \psi^n \binom{2p}{n} |n\rangle, \qquad \psi = - \tan(\theta/2)\; .
\end{equation}
Including these results in Eqs.~(\ref{S}) and (\ref{SSd}) gives a complete solution for the density matrix of the steady-state.

From this general solution it is possible to compute the expectation value of any local observable,\cite{Popkov2013a} the most important of which is the spin current $J_i$ leaving site $i$ toward site $i+1$. It is defined from the continuity equation 
\[
\frac{\ud \langle \sigma_i^z \rangle}{\ud t} = J_{i-1} - J_i
\]
where
\[
J_i = \langle \sigma_i^x \sigma_{i+1}^y - \sigma_i^y \sigma_{i+1}^x\rangle
\]
These equations are valid for $i = 2,\ldots, N-1$. 
Slightly different equations apply to the boundaries. 
In the steady state $\ud \langle \sigma_i^z \rangle/\ud t = 0$ which gives 
\[
J_1 = J_2 = \ldots = J_N := J
\]

The expectation value of an arbitrary observable $A$ may be computed as 
\[
\langle A \rangle = \tr (A \rho) = \frac{\tr(S^\dagger A S)}{\tr(S^\dagger S)}
\]
Our strategy will be to first trace over the Hilbert space and write everything in terms of expectation values on the auxiliary space. 
But note that $S$ and $S^\dagger$ will each contain an auxiliary space. So when we write $SS^\dagger$ we must double our auxiliary space. That is, we write
\[
SS^\dagger = \langle 0,0 | \Omega(p) \Omega\trans(p^*) | \psi,\psi^*\rangle
\]
where $\Omega(p)$ and $\Omega\trans(p^*)$ act on different auxiliary spaces. Moreover, $|\psi^*\rangle$ is defined as $|\psi\rangle$ in Eq.~(\ref{psi}), but with $p^*$ instead of $p$. Similarly, $\Omega\trans(p^*)$ is defined in a way similar to Eq.~(\ref{omega}): 
\[
\Omega\trans(p^*) := T_z \sigma^z + T_+ \sigma^- + T_- \sigma^+
\]
where the operators $T_a$ are defined with $p^*$ instead of $p$. Moreover, they commute with the $S_a$ since they act on different auxiliary spaces. 

Next  define 
\begin{equation}\label{Ba} 
B_a = \tr [\sigma^a \Omega(p) \Omega\trans(p^*)], \qquad a \in \{0,x,y,z\}\; .
\end{equation}
Explicitly we have 
\begin{IEEEeqnarray}{rCl}
\label{B0} B_0 &=& 2 S_z T_z + S_+ T_+ + S_- T_- \\[0.2cm]
\label{Bx} B_x &=& (S_- - S_+)T_z + S_z (T_- - T_+)\\[0.2cm]
\label{By} B_y &=& i [S_z (T_- + T_+ ) - (S_- + S_+)T_z]\\[0.2cm]
\label{Bz} B_z &=& S_+ T_+ - S_- T_- \; .
\end{IEEEeqnarray}
The spin current may then be written as 
\[
J_i = \frac{1}{Z(N)} \langle 0,0 | B_0^{i-1} [B_x, B_y] B_0^{N-i-1} |\psi,\psi^*\rangle
\]
where $Z(N)$ is the normalization constant, 
\begin{equation}\label{Z} 
Z(N) = \tr (\rho) = \langle 0,0 | B_0^N |  \psi,\psi^* \rangle\; .
\end{equation}
The explicit computation of $Z(N)$ requires constructing the matrix 
\[
(B_0)_{k,\ell} = 2|p-k|^2\delta_{k,\ell} + \ell^2 \delta_{k,\ell-1} + |2p-\ell |^2 \delta_{k,\ell+1}.
\]
We then have
\[
Z(N) = \sum\limits_{k=0}^N (B_0^N)_{0,k} \tan^{2k} (\theta/2) \left| \binom{2p}{k}\right|^2.
\]
In particular, when $\theta = 0$ one has simply $Z(N) = (B_0^N)_{0,0}$.

It can be further shown that 
\[
[B_x, B_y] = 2i (T_z - S_z) B_0
\]
and that $T_z - S_z$ commutes with $B_0$. 
This  reflects the translational symmetry of  the $J_i$ in the steady-state. 
Hence, making use of Eq.~(\ref{Sz}), we arrive at 
\begin{equation}\label{J}
J = \frac{2 \gamma}{\gamma^2 + h^2} \frac{Z(N-1)}{Z(N)}\; .
\end{equation}
This is the required formula for the steady-state magnetization flux. 
In the present model $J$ is a function of $N$, $h$, $\gamma$ and $\theta$ only. 
Eq.~(\ref{J}) must be computed for each $N$. 
Even though this may be done exactly, the formulas become extremely cumbersome for large sizes.
On the other hand, computing $J$ numerically is now a trivial task.

Also of interest is the magnon density $\langle n_i \rangle = (1+\langle \sigma_i^z \rangle)/2$. 
A calculation similar to the above leads to 
\begin{equation}\label{magnon}
\langle \sigma_i^z \rangle = \frac{\langle 0,0|B_0^{i-1} B_z B_0^{N-i}|\psi,\psi^*\rangle}{Z(N)}
\end{equation}

%
%
%
%

\section{\label{res}Results}

We now discuss the behavior of $J$ as a function of $N$, $h$, $\gamma$ and $\theta$. 
The focus will be on the case $f = 1$, for which the MPS solution is valid. 
Notwithstanding, a few words will be given about the case $f<1$. 

\begin{figure}
\centering
\includegraphics[width=0.23\textwidth]{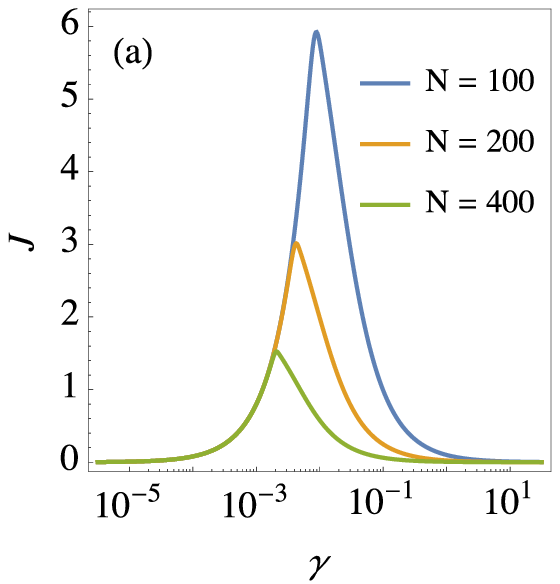}\quad 
\includegraphics[width=0.23\textwidth]{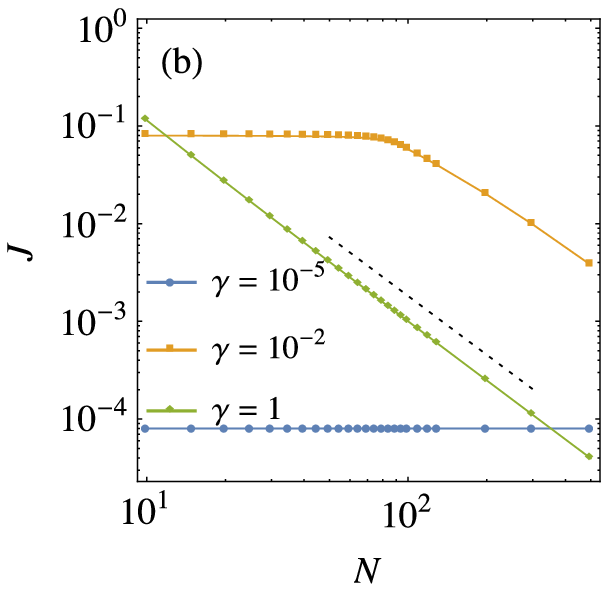}
\caption{\label{fig:h0} 
(Color Online)
Spin current $J$ for $f=1$, $\theta = 0$ and $h = 0$. 
(a) $J$~vs.~$\gamma$ for different sizes $N$.
(b) $J$~vs.~$N$ for different values of $\gamma$.
The dotted black line has slope -2.
}
\end{figure}

We begin with $\theta = 0$ and $h=0$.
The spin current as a function of $N$ and $\gamma$ is presented in Fig.~\ref{fig:h0}.
In order to interpret these results, recall  that magnons are constantly being pumped at the left source, which then  propagate through the lattice and are eventually collected in the right drain. The spin current is then simply proportional to the number of magnons being collected at the right drain. 
This number depends on two things: (i) the number of magnons being injected per unit time in the left source, which is proportional to  $\gamma$ and (ii) the magnon scattering events during the trip to the right drain. 
In standard electrical conduction (e.g.~in Drude's model), the electrons scatter with lattice imperfections or phonons. 
Since the number of  scattering agents scales proportionally to $N$ we then have a diffusive current $J \sim 1/N$.
In our case the magnons do not scatter with lattice imperfections. 
They either travel through unimpeded or they participate in 4-magnon scattering events (where 2 magnons scatter producing two new magnons in the process \footnote{The Heisenberg Hamiltonian prohibits 3 magnon events since it conserves the number of excitations. \cite{Gurevich1996} Events involving more than 4 magnons are in principle possible, but much less likely. In real systems 3-magnon events do exist as a consequence of  more complex interactions (e.g. dipolar). However, as shown in Ref.~\onlinecite{Chumak2014}, 4-magnon events are the most relevant type, at least for YIG.}).  
When $\gamma$ is sufficiently small the density of magnons in the chain is very small, thus making these events very rare. 
In this case $J$ will increase with $\gamma$ and will also be \emph{independent} of $N$; i.e., \emph{ballistic}. This is clearly observed in Fig.~\ref{fig:h0}(a), where we see that the curves for different $N$ overlap when $\gamma$ is small.
Conversely, in the high $\gamma$ limit the number of  magnons, and hence the number of scattering events, will be significant. 
In this regime it is found \cite{Popkov2013a} that $J$ is sub-diffusive, behaving as $J \sim 1/N^2$. The reason for this is that by doubling the size of the chain, we quadruple the number of four-magnon scattering events. 
As shown in Ref.~\onlinecite{Popkov2013a} the transition between the ballistic and sub-diffusive regimes occurs at 
\begin{equation}
\gamma^* \simeq \frac{1}{N}
\end{equation}
A clear example of this transition is seen in the curve for $\gamma = 10^{-2}$ in Fig.~\ref{fig:h0}(b), where the regime changes abruptly from $J\sim 1$ to $J\sim N^{-2}$ exactly  at $N = 100$.

%
%
%
%

\begin{figure}
\centering
\includegraphics[width=0.23\textwidth]{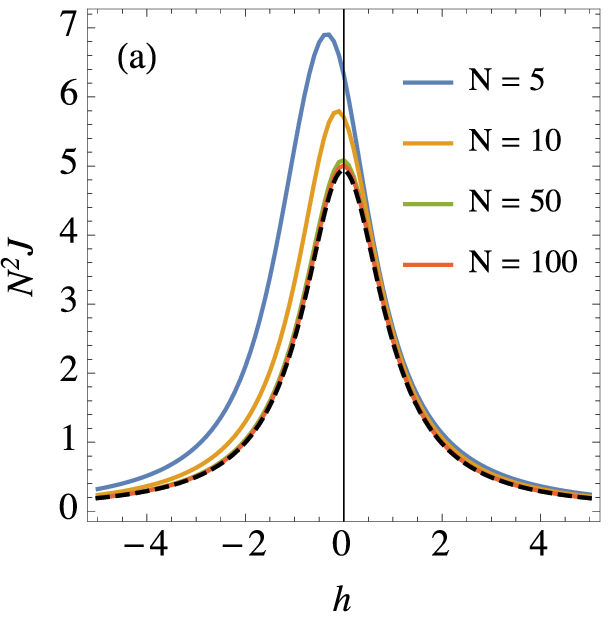}\quad 
\includegraphics[width=0.23\textwidth]{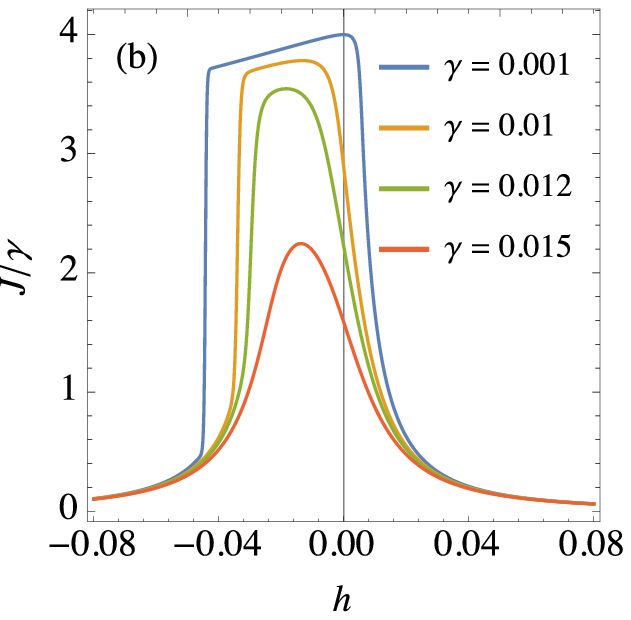}\\
\includegraphics[width=0.23\textwidth]{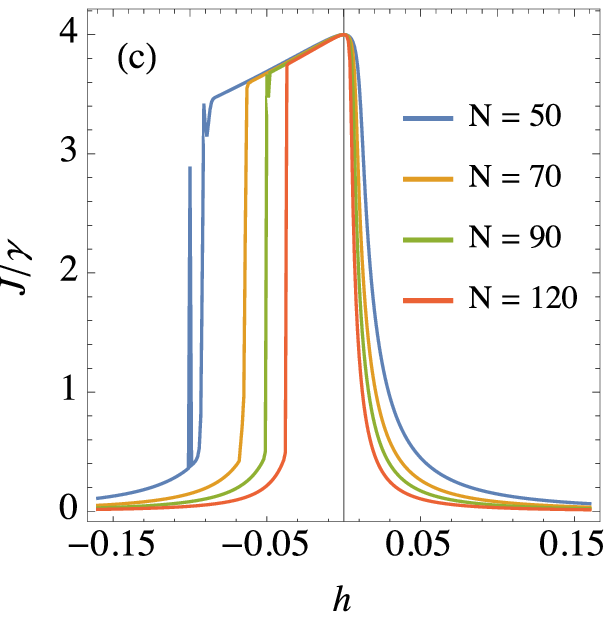}\quad 
\includegraphics[width=0.23\textwidth]{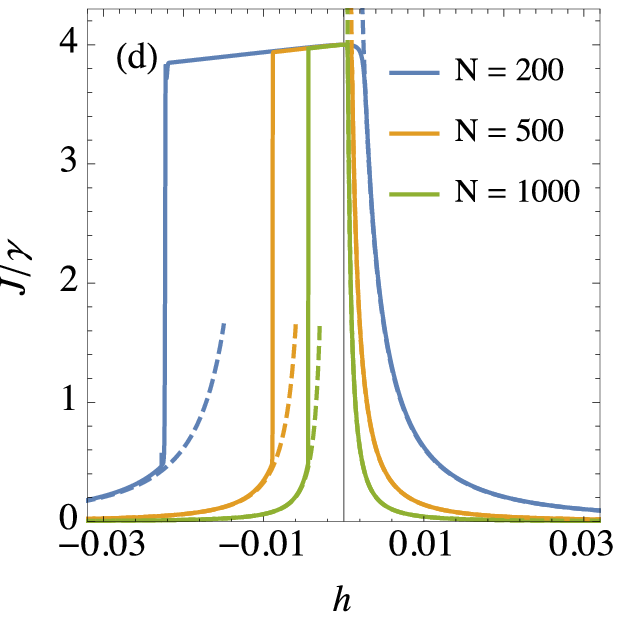}
\caption{\label{fig:h} 
(Color Online)
Spin current $J$ as a function of the boundary fields $h$ with  $f=1$ and $\theta = 0$. 
(a) $N^2 J$~vs.~$h$ for  $\gamma = 1$ and different values of $N$. 
(b) $J/\gamma$~vs.~$h$ for $N = 100$ and different values of $\gamma$ around $\gamma^* = 1/N = 0.01$. 
(c) and (d) $J/\gamma$~vs.~$h$ for  $\gamma = 10^{-5}$ and different values of $N$. 
The dashed lines in (d) correspond to Eq.~(\ref{approx}).
}
\end{figure}

Next we discuss the behavior for non-vanishing boundary fields, $h\neq 0$, still keeping $\theta = 0$. 
In Fig.~\ref{fig:h}(a) we present $N^2 J$~vs.~$h$ for $\gamma = 1$ (sub-diffusive; high magnon density).
As can be seen, even for moderately small sizes, the curves start to scale very well according to $J\sim 1/N^2$. 
In this scaling region we have found that the current is very well described by 
\begin{equation}\label{approx}
J \simeq \frac{\pi^2}{\gamma N^2} \frac{1}{1 + \frac{2h}{\gamma^2 N} + \frac{h^2}{\gamma^2}}\;,
\end{equation}
which is illustrated by the dashed line in Fig.~\ref{fig:h}(a). 
Note also that $J$ is asymmetric with respect to $h$; i.e., the spin current is rectified.\cite{Landi2014b} 

The changes which occur as we reduce $\gamma$ below $\gamma^*$ are illustrated in Fig.~\ref{fig:h}(b), where we plot  $J/\gamma$~vs.~$h$ for $N = 100$ and different values of $\gamma$. 
As can be seen, there is a drastic behavioral transition from a bell-shaped structure at $\gamma > \gamma^*$ to a \emph{plateau} at $\gamma < \gamma^*$. 
This plateau is illustrated in more detail in Figs.~\ref{fig:h}(c) and (d) for $\gamma = 10^{-5}$ and different sizes. 
As can be seen, the plateau region is asymmetric with respect to $h$ and \emph{independent of size}. 
It corresponds to the ballistic behavior of the spin current.
As the field is increased, however, one eventually observes an abrupt transition to a much lower spin current. 
For positive fields the transition is continuous whereas for negative fields it is discontinuous (strictly speaking, it is only  discontinuous  in the thermodynamic limit). 
The critical field where the plateau transition occurs is found from the simulations to be $h^* \simeq - 5/N$. 
We also call attention to the fact that outside the plateau region, $J$ is again well described by Eq.~(\ref{approx}), as illustrated by the dashed lines in Fig.~\ref{fig:h}(d). 
This  indicates that for large fields the behavior is again sub-diffusive.

%
%
%
%

\begin{figure}
\centering
\includegraphics[width=0.23\textwidth]{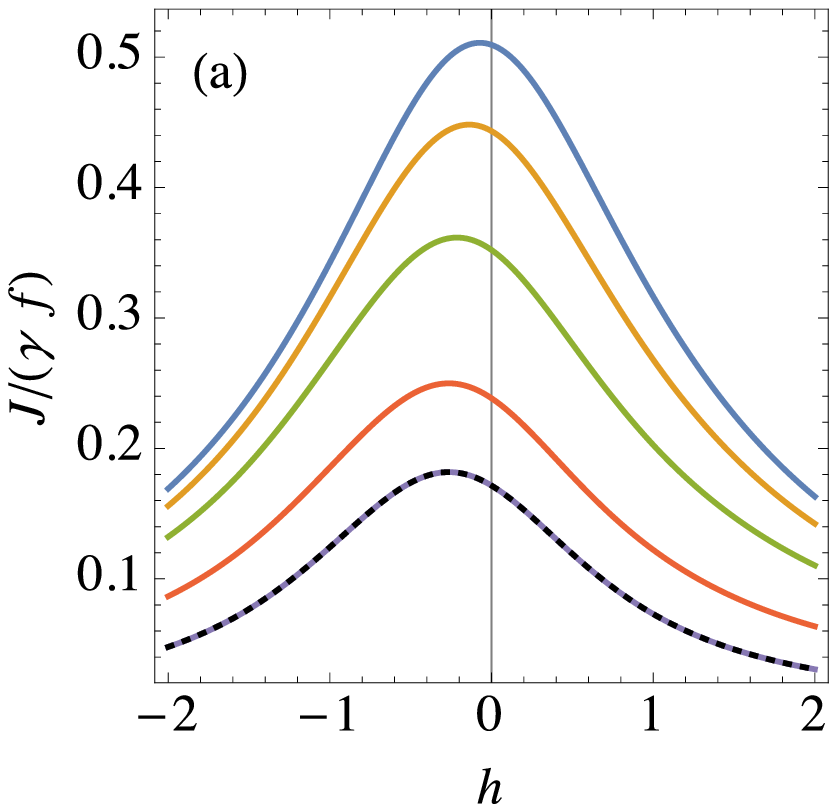}\quad 
\includegraphics[width=0.23\textwidth]{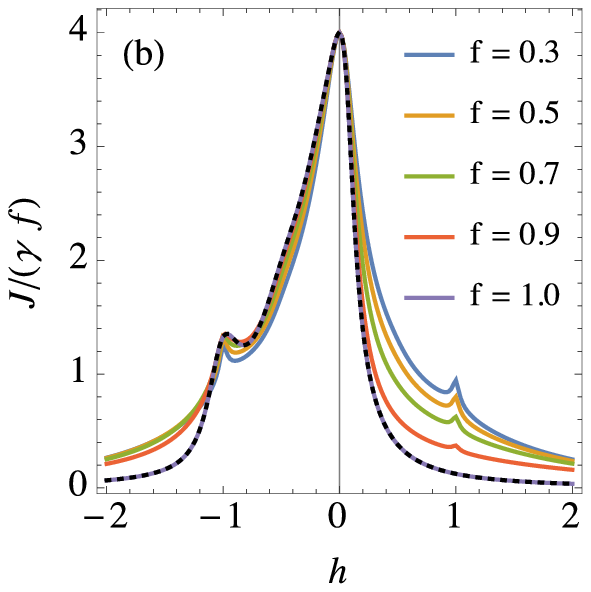}
\caption{\label{fig:f} 
(Color Online)
Spin current $J$~vs.~$h$ when $f<1$, computed using the exact diagonalization of Eq.~(\ref{lindblad}) for $N = 6$.
(a) $\gamma = 10^{-5}$.
(b) $\gamma = 1$.
The dotted black lines correspond to the MPS solution when $f = 1$.
}
\end{figure}

The results presented so far were obtained from the exact MPA steady state which is valid only at $f=1$ (zero temperature).
However, the rich behavior of the current observed for $f=1$  also survives at finite temperatures; i.e., for $f<1$. 
This can be seen in Fig.~\ref{fig:f} where we  report the current $J$~vs.~$f$ as obtained from the exact numerical diagonalization \cite{Landi2014b} of Eq.~(\ref{lindblad}) for $N = 6$. 
The current as seen from the numerics  shows basically the same features as in the MPA case: 
a bell-shaped behavior at high $\gamma$ and a sharp plateau at low $\gamma$  (for this small size the plateau is not yet completely formed). 
In Fig.~\ref{fig:f} we also plot the MPS solution when $f = 1$ to illustrate the perfect agreement between both methods.

\begin{figure}
\centering
\includegraphics[width=0.23\textwidth]{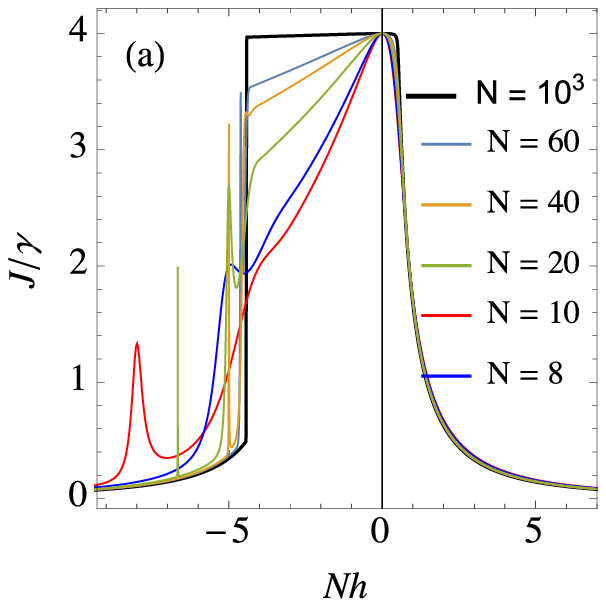}\quad 
\includegraphics[width=0.23\textwidth]{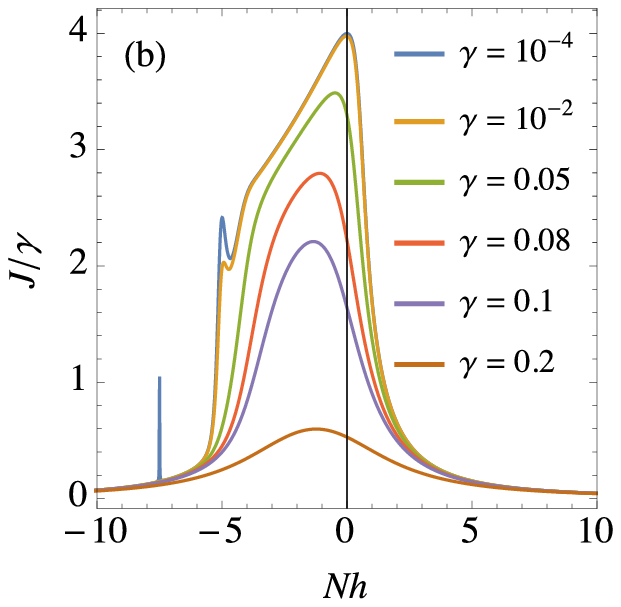}
\caption{\label{fig:small} 
(Color Online)
Small size effects in the spin current.
(a) $J/\gamma$~vs.~$Nh$ for $\gamma = 10^{-5}$ and different values of $N$. 
(b) $J/\gamma$~vs.~$Nh$ for $N = 15$ and different values of $\gamma$.
}
\end{figure}

The gradual formation of the plateau as the size of the system increases in illustrated in Fig.~\ref{fig:small}(a).
In Fig.~\ref{fig:small}(b) we show the changes which occur as one changes $\gamma$ when $N = 15$.
As can be seen in both images and in Fig.~\ref{fig:h}(c), when $N$ is small the current presents a series of irregular and sharp resonances when $h<0$, at positions which vary with $N$
(such peaks have been observed recently in Ref.~\onlinecite{Prosen2014}).
It is important to note, however, that these peaks only appear for $\gamma \leq 1/N^2$ and therefore become vanishingly small for any moderately large size. 
This can be seen, for instance, by comparing the curves with $\gamma = 10^{-4}$ and $\gamma = 10^{-2}$ in Fig.~\ref{fig:small}(b).
Both are practically identical, except for the peaks, which are only present when $\gamma = 10^{-4}$.
Note also that it follows from Eq.~(\ref{J}) that $J$ is bounded so that these cannot be delta peaks.

\begin{figure}[!h]
\centering
\includegraphics[width=0.4\textwidth]{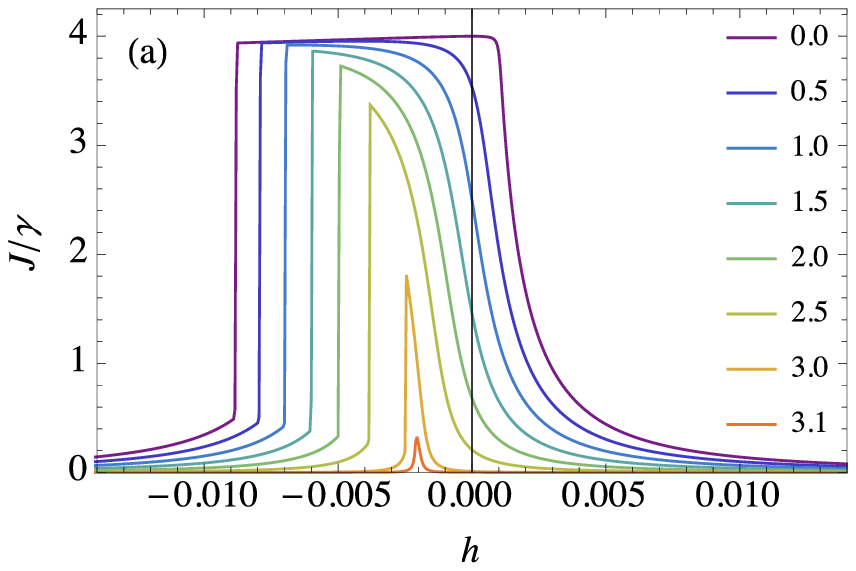}\\
\includegraphics[width=0.4\textwidth]{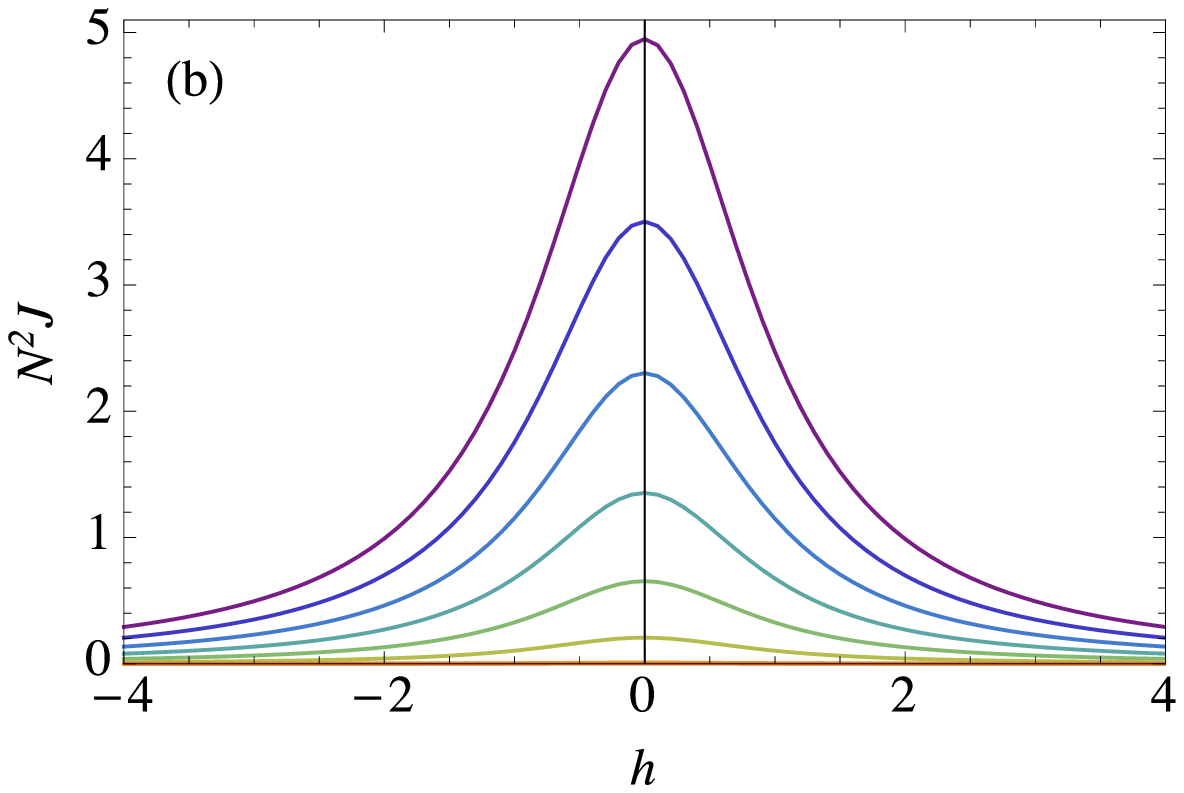}
\caption{\label{fig:theta}(Color Online) $J$ vs.~$h$ for $N = 500$ and different values of $\theta$ (as defined in Fig.~\ref{fig:drawing}).
(a) $\gamma = 10^{-4}$ and
(b) $\gamma = 1$.
} 
\end{figure}

%
%
%
%
We consider now the case with a general twisting angle $\theta \in [0,\pi]$. Fig.~\ref{fig:theta}(a) shows $J$~vs.~$h$ for different values of $\theta$ with fixed size $N = 500$ and $\gamma = 10^{-4}$. 
As expected, $J \to 0$ as $\theta\to\pi$. 
However, and remarkably, even for values of $\theta$ close to the undriven situation $\theta=\pi$, one still observes high values of $J$ for negative values of $h$, in a plateau region that shrinks as $\theta\rightarrow \pi$. 
Thus, by monitoring the twisting angle $\theta$, one can fine-tune the high current plateau width. 
For completeness, we also show the behavior for  large $\gamma$  in Fig.~\ref{fig:theta}(b).

%
%
%
%

\section{\label{sec:dis}Discussion and conclusions}

\begin{figure}[!t]
\centering
\includegraphics[width=0.23\textwidth]{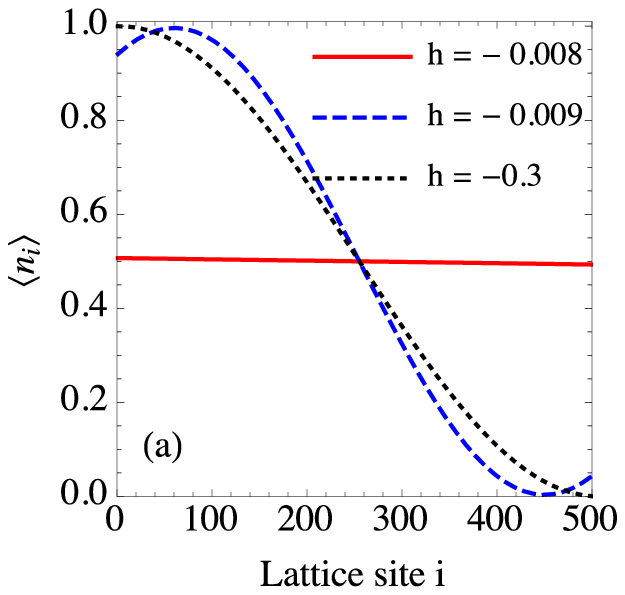}\quad 
\includegraphics[width=0.23\textwidth]{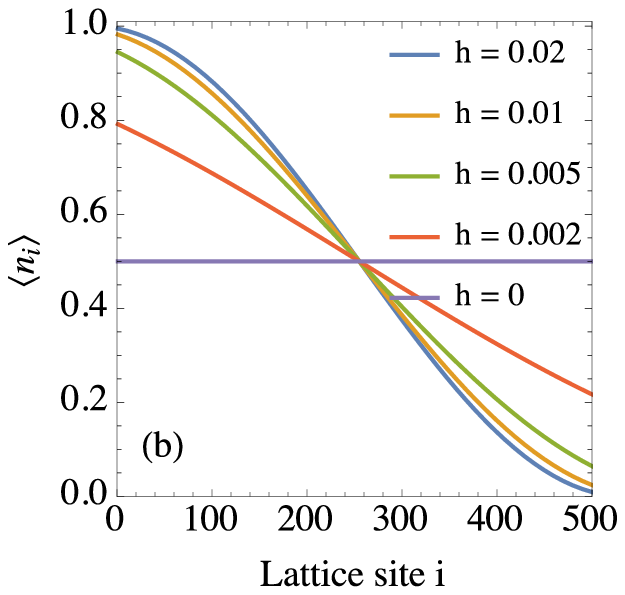}
\caption{\label{fig:density} (Color Online) 
Magnon density profile $\langle n_i \rangle = (1+\langle \sigma_i^z \rangle)/2$ for different values of $h$, with $N = 500$, $\gamma = 10^{-5}$ and $\theta = 0$.
(a) $h < 0$ near the plateau transition [cf.~Fig.~\ref{fig:h}(d)].
(b) $h > 0$.
}
\end{figure}

The  remarkable and sharp transitions observed in the spin current, from ballistic  (inside) to sub-diffusive  (outside  the plateau), as the magnitude $|h|$ of the boundary fields is increased, suggest that sufficiently high fields act as scattering barriers, impeding magnons to flow through the system, from source to drain.
This can also be seen by looking at the magnon density profile $\langle n_i \rangle = (1+\langle \sigma_i^z \rangle)/2$  plotted in Fig.~\ref{fig:density} for $N = 500$, $\gamma = 10^{-5}$ and $\theta = 0$. 
The red (solid) curve in Fig.~\ref{fig:density}(a) corresponds to the profile in the plateau (ballistic) region of Fig.~\ref{fig:h}(d). 
In this case, the distribution  is flat with $\langle n_i \rangle \simeq 1/2$, characteristic of a maximal current state. 
On the other hand, outside the plateau the profile is sine-shaped, characteristic of the sub-diffusive regime. \cite{Prosen2011b}
The transition between the two profiles is discontinuous for $h<0$ [Fig.~\ref{fig:density}(a)]  and continuous for $h>0$ [Fig.~\ref{fig:density}(b)].
Hence, we conclude that the density of magnons inside the chain may also be adjusted by changing the boundary field $h$. 
Chumak \emph{et.~al.} \cite{Chumak2014} used a similar idea to construct their magnonic logic gate. 
But in their case an additional source of magnons was responsible for changing the magnon current and the magnon density. 
Consequentially, the transition between the  on and off states was in their case quite smooth. 
Here we see an extremely abrupt transition, thus being potentially more suited for a logic gate.

In what concerns an experimental realization of the present idea, it is important  to note that  even though we studied a very specific situation, the underlying physical principles of our results are very general, being based only on the entrapment of magnons by magnetic fields. 
Hence, similar results should be obtained in different field configurations which maintain the same principles. 
Most magnonic circuits are constructed using Yttrium iron garnet  (YIG), \cite{Douglass1960,Serga2010} which is well described by the Heisenberg model, albeit with a different spin value.  
The Lindblad generators then represent microstrip antennas which are used to generate and collect magnons. \cite{Chumak2014,Serga2010}
Even though the Lindblad dissipators have been extensively used in the past to study open quantum systems, we are unaware of any papers mentioning this specific application of them as describing the injection and collection of magnons.

The energy and time units of the problem are set by the constant $\mathcal{J}$ which should appear in the first term of Eq.~(\ref{H}), but which we have throughout set as unity. 
According to Ref.~\onlinecite{Douglass1960}, $\mathcal{J} \sim 10^{-22}$ J. 
The pumping rate $\gamma$ (measured in magnons per second) should operate below the critical value $\gamma^*$ which, in the correct units, reads $\gamma^* = \mathcal{J}/\hbar N \sim 10^{12}/N$ Hz. 
This gives the optimal value of $\gamma$ below which the flux should be ballistic. 
Letting $h = \mu_B B$, where $\mu_B$ is the Bohr magnetron, we find that the  critical magnetic field $B^*$ where the plateau transition occurs is, in correct units, 
$|B^*| (\text{T}) \simeq \mathcal{J}/\mu_B N \simeq 10/N$. Hence, for any reasonable values of $N$, very small magnetic fields may suffice to induce the plateau transition. 

In summary we have studied the quantum Heisenberg chain driven by two Lindblad baths and subject to two magnetic fields acting on each boundary.  An exact solution was given in terms of matrix product states which enables one to calculate local observables for any chain size. The system is seen to undergo a discontinuous transition from ballistic to sub-diffusive spin current as a function of the field intensity. Thus, the system may function as an extremely sensitive magnonic logic gate using the boundary fields as the base.

\begin{acknowledgements}
The authors would like to acknowledge the S\~ao Paulo Funding Agency (FAPESP) and SPIDER for the financial support. 
\end{acknowledgements}

\appendix

%
%

\bibliography{/Users/gtlandi/Documents/library}

\end{document}